\documentclass[aps,pre,twocolumn,groupedaddress]{revtex4}
\usepackage{graphics}
\usepackage{graphicx}

\begin{document}


\title{Pseudochaos and low-frequency percolation scaling for turbulent diffusion in magnetized plasma}


\author{Alexander V. Milovanov}
\email[]{alexander.milovanov_at_enea.it}
\altaffiliation{Also at: Department of Space Plasma Physics, Space Research Institute, Russian Academy of Sciences, Profsoyuznaya 84/32, 117997 Moscow, Russia}

\affiliation{Associazione Euratom-ENEA sulla Fusione, Centro Ricerche Frascati, Via E. Fermi 45, C.P. 65, I-00044 Frascati, Rome, Italy}



\begin{abstract}
The basic physics properties and simplified model descriptions of the paradigmatic ``percolation" transport in low-frequency, electrostatic (anisotropic magnetic) turbulence are theoretically analyzed. The key problem being addressed is the scaling of the turbulent diffusion coefficient with the fluctuation strength in the limit of slow fluctuation frequencies (large Kubo numbers). In this limit, the transport is found to exhibit pseudochaotic, rather than simply chaotic, properties associated with the vanishing Kolmogorov-Sinai entropy and anomalously slow mixing of phase space trajectories. Based on a simple random walk model, we find the low-frequency, percolation scaling of the turbulent diffusion coefficient to be given by $D/\omega\propto Q^{2/3}$ (here $Q\gg 1$ is the Kubo number and $\omega$ is the characteristic fluctuation frequency). When the pseudochaotic property is relaxed the percolation scaling is shown to cross over to Bohm scaling. The features of turbulent transport in the pseudochaotic regime are described statistically in terms of a time fractional diffusion equation with the fractional derivative in the Caputo sense. Additional physics effects associated with finite particle inertia are considered.
\end{abstract}


\maketitle

\section{Introduction}

It is generally agreed that the presence of low-frequency, long-wavelength fluctuations in hot magnetized plasma may have a deteriorating effect on the plasma confinement leading to an anomalously high heat and energy transfer across magnetic field lines as compared to purely collisional values, a phenomenon known as anomalous or turbulent transport. While a first principles theory of turbulent transport is not at hand, the practical evaluation of the transport level is often obtained by scaling relations. An important problem investigates the scaling laws for the diffusion coefficient as a function of the fluctuation strength or the so-called Kubo number $Q \simeq u_{\bot} / \omega\xi_{\bot}$, where $u_{\bot}$ is the characteristic flow velocity, $\xi_{\bot}$ is the cross-field correlation length, and $\omega$ is the typical fluctuation frequency. 

Since the early studies of Isichenko and co-workers \cite{Kalda,Kalda2,Horton,Isi} it has been discussed by a few authors \cite{Misguich,Vlad,Vlad2,Zim00,Zim01,Veltri} that the diffusion coefficient due to turbulence exhibits a power-law dependence $D / \omega \propto Q^\gamma$, where the scaling is quasilinear-like ($\gamma = 2$) for $Q\ll 1$ and percolation-like ($\gamma = 7/10$) for $Q\gg 1$ (as distinct from Bohm scaling with $\gamma = 1$ \cite{Bohm}). It has also been discussed \cite{Vlad,PRE01,PScripta} that the percolation scaling with $\gamma = 7/10$ is however not exact and that the true value of $\gamma$ is actually smaller than (although remarkably close to) Isichenko's original estimate, $7/10$. More so, an improved percolation scaling with $\gamma = 2/3$ \cite{PRE01} has been proposed in connection with a fractional generalization \cite{Chaos,PhysicaD,Saichev} of the Fokker-Planck-Kolmogorov equation. This deviation from the earlier predicted $\gamma = 7/10$ finds support in the numerical simulation of anisotropic magnetic field turbulence \cite{PScripta,Zim01}. 

Although small, the observed discrepancy needs to be addressed. Indeed this discrepancy leads to noticeable variation in the predicted transport level (because of the large $Q\gg 1$). Apart from the numerical differences, the basic physics origin of the percolation scaling, as well as the fundamental reason for the deviation from $7/10$, has not been clearly understood.  

In this work, we expose a few crucial physics aspects behind the percolation scaling of the turbulent diffusion coefficient. We find that a mathematically consistent approach to turbulent diffusion in the limit of low frequencies (large Kubo numbers) can be obtained within the concept of {\it pseudochaos} (random non-chaotic dynamics with zero Lyapunov exponents) \cite{Report,JMPB,PD2004}. Our analysis displays a few characteristic features of pseudochaos differentiating it from the more intuitive, chaotic behavior. We confirm the percolation scaling with $\gamma = 2/3$, basing our considerations on a simple random-walk model in fractal geometry. Yet, a slightly smaller value $\gamma\simeq 0.66$ could be advocated involving subtleties of the random walks at percolation. 

More so, we demonstrate that the diffusion on percolation systems is described by a non-Markovian diffusion equation, with the non-Markovian property expressible in terms of a fractional time derivative in the Caputo sense \cite{Podlubny}. Here the stress is on ``Caputo" and on subtleties of definition of the fractional derivative operator. Being conceptually very similar, this definition is somewhat different from the definition in the Riemann-Liouville sense \cite{Oldham}, of wider use in the applications. Indeed fractional diffusion equations with the Riemann-Liouville derivative have, in the various contexts, been studied and discussed in the literature (Refs. \cite{Klafter,Sokolov,Rest,Uchaikin} for reviews).  

The paper is organized as follows. The concepts of chaos, pseudochaos, and resonances in Hamiltonian dynamics are discussed first (Sec. II), followed by a derivation of the percolation scaling from a random walk model in fractal geometry (Sec. III). Fractional derivative equations are obtained in Sec. IV in the framework of generalized memory function formalism. Next issues related to particle inertia are considered (Sec. V). We summarize our findings in Sec. VI. Applications of this study pertain to both geo-space \cite{Galeev,UFN} and fusion \cite{Zonca,Zonca2} plasma, and by mathematical analogy to problems outside the plasma physics. 

\section{Hamiltonian, resonances, and pseudochaos}

To begin, we formulate the transport problem as the Hamiltonian problem      
\begin{equation}
\frac{dx}{dt} = \frac{\partial H(x,y,t)}{\partial y}, \ \ \ \frac{dy}{dt} = -\frac{\partial H(x,y,t)}{\partial x}, \label{1} 
\end{equation} 
where $x$ and $y$ are coordinates in the plane perpendicular to the magnetic field, and $H(x,y,t)$ is a time-depending Hamiltonian. 

\subsection{Hamiltonian problem - examples}

The physics included in Eqs.~(\ref{1}) encompasses, beside others, the following model realizations: 

(i) Field-line diffusion by anisotropic magnetic turbulence. Early work on this problem is due to Rosenbluth {\it et al.} \cite{Taylor}. The magnetic field model is set as a sum of a constant homogeneous background field ${\bf B}_0 = B_0 \hat{\bf z}$ in a straight-cylinder geometry (here $\hat{\bf z}$ is a unit vector in the axial direction) and a static transverse magnetic perturbation, $\delta {\bf B}_\bot ({\bf r}_\bot, z)$, small compared to the background field: $\delta B_\bot / B_0 \ll 1$. The equations for the magnetic field lines are cast in the Hamiltonian form by defining $\delta {\bf B}_\bot = \nabla A_\| ({\bf r}_\bot, z)\times\hat{\bf z}$, where ${\bf r}_\bot = (x,y)$ is the position vector in the perpendicular plane, ${\bf r}_\bot\cdot\hat{\bf z}=0$, $\nabla = \partial / \partial {\bf r}_\bot$, $H({\bf r}_\bot,z) = A_\| ({\bf r}_\bot, z) / B_0$ is the Hamiltonian, and the axial coordinate $z$ is considered as ``time." The flow velocity is given by ${\bf u}_\bot = \nabla H ({\bf r}_\bot,z)\times\hat{\bf z} = \delta {\bf B}_\bot / B_0$, making it possible to evaluate $Q\simeq u_\bot/\omega \xi_\bot \simeq (1 / 2\pi) (\delta B_\bot / B_0) (\xi_\| / \xi_\bot)$, where $\delta B_\bot$ is the rms magnetic fluctuation. In writing the last form we took into account that the ``frequency" of the field variation due to the $z$ dependence is $\omega \simeq 2\pi / \xi_\|$, where $\xi_\|$ is the typical parallel correlation length. It is noted that $Q\gg 1$ implies $\xi_\| / \xi_\bot \gg 1$ with a large margin, and hence, it is required the magnetic turbulence be strongly anisotropic.   

(ii) Diffusion of guiding centers by the ${\bf E}_\bot\times {\bf B}$ drift. This problem refers to the well-known problem of particle diffusion by low-frequency, electrostatic, microturbulence \cite{Misguich,Naulin,Horton2}. This is usually one of the most important transport problems in a fusion plasma. The Hamiltonian is written as $H({\bf r}_\bot,t) = - (1 / B_0)\Phi({\bf r}_\bot,t)$, where $\Phi({\bf r}_\bot,t)$ is the electrostatic potential in the perpendicular plane, such that ${\bf E}_\bot ({\bf r}_\bot,t) = -\nabla\Phi ({\bf r}_\bot,t)$, and $B_0$ is the amplitude of a confining, uniform magnetic field. The control parameter is defined by $Q\simeq E_\bot / \omega \xi_\bot B_0$, where $E_\bot$ is the rms electric field, and $\omega$ is the typical fluctuation frequency.

(iii) Diffusion of guiding centers by the $\nabla B$ drift. Consider the magnetic field in the $z$ direction and suppose that it is spatially nonuniform in the perpendicular plane in accordance with ${\bf B} = B ({\bf r}_\bot, t)\hat{\bf z}$, $B ({\bf r}_\bot, t) = B_0 + \delta B ({\bf r}_\bot, t)$, and $\delta B / B_0 \ll 1$. The time dependence in $\delta B ({\bf r}_\bot, t)$ is assumed to be very slow, so that the corresponding variation frequency is, with a large margin, small compared to a characteristic cyclotron frequency, i.e., $\omega / \omega_c \ll 1$. It is understood that the time varying magnetic perturbation generates a time and spatially varying electric field because of Faraday's law. A consistent description of the guiding center motion, then, will require that the ${\bf E}_\bot\times {\bf B}$ and polarization drifts are taken into consideration. A simplified yet relevant model is obtained for situations in which the spatial inhomogeneities of the magnetic field are attributed a major role in determining the characteristic transport properties of the considered system as compared to the explicit time dependencies of the magnetic fluctuation and inductive electric fields. For instance, such model conditions can occur for the transport of solar wind-like plasma inside the magnetosphere as a consequence of the internal fine structure of the Earth's magnetopause \cite{Galeev}. In this spirit, if we assume that the $\nabla B$ drift velocity is the dominating velocity in the Alfv\'en approximation, and neglect the higher-order corrections due to the inductive electric field, we can write ${\bf u}_\bot \simeq \mp \mu_B (\hat{\bf z}\times\nabla B)/qB_0$, where the upper sign corresponds to a negative charge and we have, following Refs. \cite{Bitten,Jeffrey}, introduced the first adiabatic invariant $\mu_B = mv_\bot^2 / 2B$ (here $v_\bot$ is the perpendicular particle velocity, $m$ is the particle mass, and other notations are standard). In order for the drift approximation be valid, we also need to require $(v_\bot / \omega_c) |\nabla B / B_0| \ll 1$. We note in passing that the $\nabla B$ drift is a finite Larmor radius effect. The Hamiltonian is defined by $H({\bf r}_\bot, t) = \pm (\mu_B / qB_0) \delta B({\bf r}_\bot, t)$. This model Hamiltonian isolates the effect of the magnetic inhomogeneity. It includes the magnetic gradients slowly changing with time. Effects due to the time varying electric field, left apart in the present model, will be considered in more detail in Sec. V.  

\subsection{Equipotentials - fractal structure}

At a given time $t$ the equipotentials of the Hamiltonian are defined by $y = y (x, H, t={\rm const})$. By standard rules the area embraced by a closed equipotential is obtained as $I = \oint y dx$ (this being the action of the Hamiltonian system), while the frequency of the orbital motion is given by $\Omega (I) = dH (I) / dI \simeq |\nabla H ({\bf r}_\bot,t)| /|\nabla I ({\bf r}_\bot,t)|\simeq u_\bot / \ell$, where $\ell \simeq |\nabla I ({\bf r}_\bot,t)|$ is the circumference of the equipotential contour, and $t$ is kept as a fixed parameter. Assuming circular equipotentials one writes $I\propto {r}_\bot^2$, $\ell \propto {r}_\bot$, and $I\propto \ell ^2$. In the case of strongly shaped equipotential contours, however, the latter scaling relations may not hold true. Of particular interest are the equipotentials characterized by self-similar geometry in a broad range of spatial scales best described as fractals \cite{Feder}. The implication of such equipotentials in diffusion by low-frequency turbulence is addressed shortly. For fractal equipotential contours the simple scaling $I\propto \ell ^2$ is generalized to $I\propto \ell ^{2/d_h}$ (the so-called ``area-perimeter relation" \cite{LeMehaute}) where $d_h$ is the fractal dimension of the equipotential line and we have suppressed the normalization parameter for simplicity. The fractal dimension $d_h$ is defined by the scaling $\ell \propto {r}_\bot^{d_h}$ (instead of $\ell \propto r_\bot$ in non-fractal geometry). The $d_h$ values are generally fractional larger than 1. Eliminating $\ell$ by means of the area-perimeter relation one obtains $\Omega (I) \propto u_\bot / I^{d_h / 2}$.             

\subsection{Hamiltonian - Fourier representation}

In the discussion so far we have not made any specific assumptions about the structure of the $H$ field in space and time. We now need to be more detailed in connection with the notion of ``turbulence." It is convenient to think of turbulence as of collection of plane waves, with amplitudes $b_{n,\bf k}$ and random phases $\phi_{n,\bf k}$, making it possible to expand the Hamiltonian into the Fourier series to give (with ${\bf k}$ being the wave vector in the perpendicular plane)    
\begin{equation}
H({\bf r}_\bot,t) \simeq \sum_n \sum_{\bf k} b_{n,\bf k} \cos({\bf k}\cdot {\bf r}_\bot - \omega_n t + \phi_{n,\bf k}). \label{5} 
\end{equation} 
Next, it is generally agreed that there is a broad, isotropic wave vector spectrum characterized by a power-law. In 2D geometry we can write  
\begin{equation}
P(k)\simeq 2\pi k \sum_n b_{n,\bf k}^2 \propto k^{-\alpha}. \label{6} 
\end{equation}
Actually a more precise way of looking at the spectral characteristics of the turbulence would be to introduce a fully 3D spectrum, squeezed to a ``pancake" in the perpendicular plane. Such spectra are considered in Ref. \cite{Zim00}. 

The next point to be addressed is the $\alpha$ value in Eq.~(\ref{6}). For the present analysis, we take $1\leq\alpha\leq 3$. Together with the inverse power-law energy density distribution, the latter condition implies that the fluctuations are {\it self-similar}, in the sense of Refs. \cite{Feder,LeMehaute}. Indeed this property of self-similaraty finds support in the many different experiments \cite{Burlaga86,JGR96,Borovsky,Carreras,Zaslav,UFN}. From the standpoint of formal treatment, the fluctuations being self-similar pave the way for the application of fractal geometry as proposed in the model below. 

To match with the physics natural limitations, the spectrum in Eq.~(\ref{6}) is characterized by finite cutoffs on both sides, i.e., $a \ll 2\pi /k \ll \xi_\bot$. Without loss in generality, we shall assume that all fields are smooth at length scales shorter than $a$ and that there is a transition to a statistically homogeneous distribution of the fluctuations at length scales longer than $\xi_\bot$. 

Consistent with the above assumptions regarding the guiding center approximation, we shall also assume that $a$ is large compared with the typical scales of the particle gyro motion. When the particle gyro radius is not small enough to match with this condition, then the particle motion can be considerably modified by the multi-scale geometry of the turbulent field. In some cases the particle trajectory can be extremely complicated. One example of this complex motion is the case of ``strange diffusion," discussed in Refs. \cite{PScripta,UFN}. We also note that, when $\xi_\bot / a \rightarrow \infty$, the expansion in Eq.~(\ref{5}) reproduces the so-called fractional Brownian function, which offers a particularly clear example of fractal behavior \cite{Feder,Siam,Affine}.  
 
\subsection{Percolation property and resonances}
 
Because of the sign symmetry of the Hamiltonian in Eq.~(\ref{5}), the zero-set $H ({\bf r}_\bot, t={\rm const}) = 0$ contains a percolating equipotential line \cite{Isi,PRE00}, which, by its geometric meaning, stretches the entire system. The percolating line is the channel through which turbulent transport penetrates to the large scales \cite{Isi}. One way to describe the transport is to directly link it to the complex folding of the percolating line in the real space and in this fashion to predict the anomalous scaling laws for the turbulent diffusion coefficient as suggested in the early work \cite{Kalda}. This approach, however, leads to some difficulties associated with the fractal dimension of the transport process as it basically ignores the dynamical features of the transport \cite{PRE01}. A correct way to deal with the properties of turbulent diffusion is to account for resonances between the orbital motions of the particles and the time variation of the Hamiltonian. As is well known, when phase space trajectories are subject to local instability, the matching fluctuation and orbital frequencies or the matching harmonics of these favor departure from the exact periodic, energy conserving motion thus giving rise to transport phenomena in phase space (Refs. \cite{Zaslavsky,ZaslavskyUFN,Sagdeev} for the full discussion). In the limit of very low frequencies, the conditions for a resonance require the corresponding excursion periods to diverge. Mathematically, this can be satisfied for fractal equipotential contours, characterized by the diverging lengths due to the structure on many scales. Analysis of Ref. \cite{PRE00} have shown that the equipotentials of the $H$ field in Eq.~(\ref{5}) occurring in close proximity to the percolating line are indeed described by the fractal structure in the limit $\xi_\bot / a \rightarrow \infty$ and their fractal dimension was found to be $d_h = 1.32\pm 0.01$. This fractal dimension is close to the value $d_h = 1.37\pm 0.03$, obtained in Ref. \cite{Aharony}, and is clearly smaller than the hull exponent, $d_h = 1.75$, used by Isichenko and co-workers \cite{Kalda,Kalda2,Isi}. Anticipating our result in Sec. III, here we mention that the ``percolation" scaling of the turbulent diffusion coefficient won't however involve the $d_h$ parameter directly, in contrast with the conclusion of Refs. \cite{Kalda,Kalda2,Isi}.   

\subsection{From the percolation property to pseudochaos}

When the fluctuations are very slow, we expect the resonance conditions to be naturally satisfied in vicinity of the percolating line, provided that the number of modes in Eq.~(\ref{5}) is large enough. As is already mentioned in the above discussion, this line, being strongly shaped, is characterized by the fractal geometry in the limit $\xi_\bot / a \rightarrow \infty$. Fractality, in its turn, guarantees the {\it existence} of resonant orbits for $\omega_n\rightarrow 0$. It is noted that, when $\omega_n\rightarrow 0$, the resonances strongly overlap in a very narrow layer enveloping the percolating line. This observation actually refers to the general properties of Hamiltonian dynamics near the separatrices, discussed in Ref. \cite{Sagdeev}. It is the overlap of resonances that introduces instability into the motion. If one defines the characteristic frequency, $\omega$, one finds the width of the resonance layer to be $\delta\omega\propto\sqrt{\omega}$ (as compared to the distance between the resonances $\Delta\omega\propto\omega$). Note that $\delta\omega\gg\Delta\omega$ for $\omega\rightarrow 0$, and that the width of the resonance layer shrinks to zero (as $\sqrt{\omega}$), while the density of the resonances in the layer, defined as inverse distance $\Delta\omega$, diverges as $1/\omega$. Inside the layer, the dynamics are basically random because of the many resonances present \cite{ZaslavskyUFN,Sagdeev}. On the other hand, the concentration of the resonance properties on a fractal structure, supported by $\xi_\bot / a \rightarrow \infty$, strongly reduces the phase space available for the random motions. As a consequence, the chaotic trajectories separate anomalously slowly. ``Anomalously slowly" means sub-exponential for $\omega\rightarrow 0$ (see the discussion below), which is equivalent to saying that the Lyapunov exponents, thought of as functions of position in phase space, vanish at percolation. In view of the vanishing $\delta\omega\rightarrow 0$ the phenomenon can be envisaged as ``stickiness" to the percolating line. More so, the vanishing Lyapunov exponents, combined with the inherently random character of the motion, indicate that the dynamics are {\it pseudochaotic}, in the sense of Refs. \cite{Report,JMPB,PD2004}, instead of being just {\it chaotic}. This is the key point. Based on the notion of pseudochaos \cite{Report}, one almost straightforwardly derives the ``percolation" scaling of the turbulent diffusion coefficient in the limit of slow fluctuation frequencies.

\subsection{$K$ parameter}

In order to make these intuitive arguments more precise, we proceed as follows. In the basic theory of Hamiltonian chaos one writes the number of overlapping resonances as $\sqrt{K}$, where the $K$ value (the so-called nonlinearity parameter) is obtained from \cite{Zaslavsky,ZaslavskyUFN,Sagdeev}           
\begin{equation}
K = \frac{I}{\omega}\left|\frac{d\Omega (I)}{dI}\right|. \label{4} 
\end{equation} 
Consistent with the above orderings, $\sqrt{K}\propto 1 / \sqrt{\omega} \gg 1$ in the resonance layer. Also, if we denote the Lyapunov exponents of the Hamiltonian system in Eqs.~(\ref{1}) by $\pm \sigma$, we can order $|\sigma|\simeq h_{\rm KS}\simeq 1/ \tau_c$. Here $h_{\rm KS}$ is the Kolmogorov-Sinai entropy and $\tau_c$ is the characteristic mixing time. For $K\gg 1$, the mixing time is written as $\tau_c\simeq 1 / 2\omega\ln K$ (Refs. \cite{Zaslavsky,Sagdeev} for the full discussion). When $\omega\rightarrow 0$, the $K$ parameter goes to infinity as $1/\omega$, while the mixing time diverges as $\tau_c\propto -1 / 2\omega\ln\omega\rightarrow\infty$. Accordingly, $|\sigma|\simeq h_{\rm KS}\simeq -2\omega\ln\omega\rightarrow 0$ at percolation. Thus, while the dynamics are intrinsically random due to the many overlapping resonances, the available phase space is insufficient to host the chaos. Indeed the mixing happens to be sub-exponential permitting the system exhibit only a pseudochaotic behavior.    

Let us now rewrite Eq.~(\ref{4}) in a more insightful form. Substituting $\Omega (I) \propto u_\bot / I^{d_h / 2}$ and performing the trivial differentiation over $I$ leads to $K \propto u_\bot / \omega\ell$, where the area-perimeter relation $I\propto \ell ^{2/d_h}$ has been applied and we have omitted the numerical constant factor for simplicity. For the fractal equipotential contours near the percolating line, the circumference $\ell \propto \xi_\bot^{d_h}$, making it possible to evaluate $K \propto u_\bot / \omega\xi_\bot^{d_h}$. Remembering that the Kubo number $Q\propto u_\bot / \omega\xi_\bot$ it is found that
\begin{equation}
Q/K \propto \xi_\bot^{d_h - 1}. \label{5+} 
\end{equation} 
Hence, in the pseudochaotic regime, $Q\gg K$ due to fractality ($\xi_\bot / a \gg 1$, $d_h > 1$). When $d_h \rightarrow 1$, the $Q$ and $K$ parameters have the same order. 

\subsection{Kolmogorov-Sinai entropy}

Eliminating $K$ by means of Eq.~(\ref{5+}) we can also express the Kolmogorov-Sinai entropy as a function of the Kubo number as $h_{\rm KS}\simeq 2\omega\ln Q - 2\omega (d_h-1)\ln\xi_\bot -2\omega\ln C$, where the last term in the sequence comes from the (suppressed) normalization constant factor in the above scaling relation. It is noted that, to the leading order, $h_{\rm KS}$ is a logarithmic function of $Q$. More generally, this logarithmic behavior can be slowed down by $d_h$ nonlinearly depending on the Kubo number. The implication is that fractal geometries of the contours on which the resonances occur can differ with the control parameter. This effect will be discussed elsewhere \cite{Unpublished}. Our result for $h_{\rm KS}$ is in contrast with Isichenko's finding $h_{\rm KS}\simeq \omega Q^{1/2} \ln Q$ for $Q\gg 1$ (Eq.~(4.55) in Ref. \cite{Isi}) and the square-root-like scaling versus $Q$, argued to be ``universal." Indeed it is found in direct numerical computations \cite{Unpublished,Zimbardo} that the square-root-like scaling of the entropy is not reproduced and that the behavior is actually slower than a power-law, tending to saturation. We interpret these results as consistent with implications of the slow mixing and pseudochaos.

\subsection{Summary}

The end result of the above reasoning is that turbulent diffusion in the limit of very low frequencies is characterized by the random dynamics squeezed to a subset of phase space with fractal geometry. In this regime $Q\gg K \gg 1$, whereas for the classical chaotic behavior \cite{Zaslavsky,Sagdeev}, characterized by the wide domains of random motion, $Q\simeq K\gg 1$. Overall, the dynamics bear signatures enabling to associate them with {\it pseudochaos} (random non-chaotic dynamics with zero Lyapunov exponents) \cite{Report}. 
 
\section{Percolation scaling for turbulent diffusion}

We are now in position to obtain the scaling law for the diffusion coefficient in the parameter range of slow fluctuation frequencies (large Kubo numbers). The calculation is based on the general scaling form  
\begin{equation}
\langle{\bf r}_\bot^2 (t)\rangle = \xi_\bot^{2} (t/\tau_*) f (t/\tau_*), \label{9} 
\end{equation}
where $\tau_*$ is the characteristic diffusion time to the distance $\xi_\bot$, and $f$ is a scaling function, such that $f(\infty) = {\rm const.}$ The form in Eq.~(\ref{9}) is similar to that considered by Gefen {\it et al.} \cite{Gefen} for anomalous diffusion on percolation clusters (in their model $\tau_*\propto\xi_\bot^{2+\theta}$, where $\theta$ is the index of anomalous diffusion), and earlier by Straley \cite{Straley}.   

Based on the above discussion, we assume the $f$ function in Eq.~(\ref{9}) has already reached its asymptotic value after the characteristic mixing time $t\simeq\tau_c$. That is, after this time, the dispersion in Eq.~(\ref{9}) is linear. On shorter time scales, the dispersion deviates from linear because of the concentration of the transport process on a fractal geometry. Since the dynamics are nevertheless random, we can envisage them as a random walk process on a fractal cluster at percolation. This process has been discussed as a simple model process for diffusion in disordered systems (``the ant in the fractal labyrinth") \cite{Gennes,Havlin}. In those models a particle (random walker) is assumed to hop in random manner between nearest neighbor sites of a fractal lattice moving a step of length $a$ each $a/u_\bot$ seconds. In more advanced models the condition that the steps occur at fixed time intervals is relaxed \cite{CTRW}. Note that, in all these models, the fractal lattice is assumed to be stationary. In turbulent diffusion, this assumption holds true for only short time lags $t\lesssim\tau_c\ll 1/\omega$. At these short times, we can require the $f$ function in Eq.~(\ref{9}) correspond with the scaling function in the simple random walk model \cite{Gefen}: $f(s)\propto s^{-\mu/\nu(2+\theta)}$. Here $\mu$ and $\nu$ are the percolation indices, whose definitions are explained in major reviews \cite{Isi,Stauffer,Naka}. It is remarked that $\tau_c$ has the sense of the characteristic cross-over time scale where the $f$ function changes from the initial-time power-law behavior to the asymptotic constant value. The diffusion coefficient is evaluated by noting that $f(\infty)\simeq (\tau_c / \tau_*)^{-\mu/\nu(2+\theta)}$, then making use of Eq.~(\ref{9}) to obtain          
\begin{equation}
D\simeq (\xi_\bot^{2}/2\tau_*) (\tau_c / \tau_*)^{-\mu/\nu(2+\theta)}. \label{10} 
\end{equation}
Focusing on the random walks at percolation we somehow expect the fractal range be wide enough, ideally $\xi_\bot / a \rightarrow\infty$. On the other hand, the slowness of the fluctuations implies that $Q\simeq u_\bot / \omega\xi_\bot\rightarrow\infty$ for $\omega\rightarrow 0$. For the purpose of formal orderings, we require $\xi_\bot / a\rightarrow\infty$ diverge faster than $Q\rightarrow\infty$. That is, given a finite fluctuation frequency, we set $\xi_\bot / a \gg u_\bot / \omega\xi_\bot$. The implication is that the fluctuations should not be too slow for the actual (finite) size of the system, otherwise the dynamics retain a deterministic character. If we define $\tau_*\simeq \xi_\bot ^2 /a u_\bot$, from the last inequality we also get $\omega\tau_* \gg 1$ showing that the diffusion time to the distance $\xi_\bot$ must be large compared to the period of the field. Because of this, we expect fractal correlations to decay before the particles have crossed the entire system. Thus, in our model, the ``anomalous" scaling $\tau_*\propto \xi_\bot^{2+\theta}$ \cite{Gefen} is invalidated in the limit $\xi_\bot\rightarrow\infty$ as a consequence of the time varying Hamiltonian. Instead, the conventional scaling law for diffusion, $\tau_*\propto \xi_\bot^{2}$, applies. Remembering that $\tau_c\simeq 1 / 2\omega\ln K$, we can cast the diffusion coefficient in Eq.~(\ref{10}) in the form 
\begin{equation}
D\simeq 2^{-\gamma} (\ln K)^{1-\gamma} (a/\xi_\bot)^\gamma \omega \xi_\bot^2 Q^\gamma, \label{11} 
\end{equation}
where the exponent $\gamma$ is defined as
\begin{equation}
\gamma = 1-\mu/\nu(2+\theta). \label{13} 
\end{equation}
This is the desired result. We have $D / \omega \propto Q^\gamma$, where $\gamma$ depends on the percolation indices, but the expression is different from the one obtained by Isichenko \cite{Isi}. Note that $D / \omega$ is not a simple scaling function of $\omega$ (because of the $K$ parameter varying with frequency). The correction is expressible as a power of the logarithmic dependence $\propto (-\ln\omega)^{1-\gamma}$, where $\omega\rightarrow 0$.   

\subsection{General expression}

It is convenient to express $\gamma$ in terms of two parameters only, the index of anomalous diffusion $\theta = (\mu - \beta) / \nu$ and the Hausdorff dimension of the infinite percolation cluster, $d_f = d - \beta / \nu$ \cite{Stauffer}. Here $d$ is the topological (integer) dimension of the ambient space. Remembering that the considered system of equations, Eqs.~(\ref{1}), is 2D, we have $d=2$. Nevertheless, the above relation, which is called the hyperscaling relation, holds in all $d\geq 2$ \cite{Naka}. In the physics context discussed in this work the Hausdorff dimension $d_f$ measures the subset of phase space occupied by the random motions. This dimension is to be distinguished from the fractal dimension of the equipotential lines, $d_h$. The implication is that the particles (random walkers) do not follow the equipotential lines exactly because of the fine structure of the resonance layer. When $d_f$ is smaller than $d$, the random motions are, by definition, not space-filling. Indeed they fill only a fractal subset of phase space, characterized by $d_f = d - \beta / \nu$. Eliminating the percolation indices in Eq.~(\ref{13}) one gets
\begin{equation}
\gamma = (2-d+d_f) / (2+\theta). \label{14} 
\end{equation}
To take the non-fractal limit in Eq.~(\ref{14}), one sets the Hausdorff dimension equal to the ambient dimension, and the index of anomalous diffusion equal to zero. That is, $d_f = d$ and $\theta = 0$. As a consequence, Bohm scaling with $\gamma = 1$ is recovered from Eq.~(\ref{14}). This limit, because of the space-filling property, corresponds with the assumptions of classical chaotic behavior. Next we focus on fractal (not space-filling) case: $d_f < d$. Setting $d=2$ in Eq.~(\ref{14}) one obtains
\begin{equation}
\gamma = d_f / (2+\theta) = d_s / 2. \label{14+} 
\end{equation}
Here $d_s = 2d_f / (2+\theta)$ is the so-called spectral fractal dimension \cite{Havlin,Naka}, which has come into physics as the density-of-state exponent for vibrational excitations of fractal networks \cite{AO}. It is noted that $d_s \leq d_f$ (because of $\theta \geq 0$). The spectral dimension has the sense of effective, fractional number of degrees of freedom in fractal geometry \cite{Bouchaud}. We draw attention to the fact that this fractional number is not determined by $d_f$ only, but also requires $\theta$. In general, $d_f$ and $\theta$ are two independent parameters, whose tradeoff defines $d_s$ [and hence the $\gamma$ value, according to Eq.~(\ref{14+})]. The situation is somewhat simpler for percolation, as we now proceed to show. 

\subsection{Numerical estimate}

In recent years there has been much excitement about the Alexander-Orbach (AO) conjecture \cite{AO} that $d_s = 4/3$ for $d\geq 2$ at percolation. The AO conjecture is exact in the high dimensions $d\geq 6$ where it holds as a mean-field result \cite{Coniglio}. In dimensions lower than these, the AO conjecture proves to be not exact, being, nevertheless, a remarkably accurate estimate of the $d_s$ value. Based on the available numerical and theoretical predictions, the true value of the spectral dimension is believed to be slightly {\it smaller} than 4/3. The actual discrepancies lie within half a percent. See Refs. \cite{Naka,Havlin,Havlin2}. As an example, consider the analytic prediction $d_s = 1.327\pm 0.001$ for $d<6$, suggested in Ref. \cite{PRE97}. If, in Eq.~(\ref{14+}), we apply the mean-field estimate $d_s = 4/3$ thus ignoring the above small discrepancy we get $\gamma = 2/3$. This value coincides with our previous finding (Eq. (28) in Ref. \cite{PRE01}, where the Hurst exponent $H=1/2$ for diffusion). If one is a purist and wants to account for the departure from mean-field properties, the procedure is to expedite the spectral dimension $d_s\approx 1.327$ \cite{PRE97} or the like \cite{Stat} to find $\gamma\simeq 0.66$, at little consequence for the rest of the analysis. The basic phenomenon is contained in Eq.~(\ref{13}). 

\subsection{Summary}

To this end, our final result for the percolation scaling is: $D / \omega \propto Q^{2/3}$, where $Q\gg K \gg 1$, and we have omitted the logarithmic correction factor for simplicity. When the random motions fill the ambient space ($d_f = d_s = d$), Bohm scaling with $\gamma = 1$ is recovered. A deviation from Bohm scaling associated with a fractional $\gamma$ smaller than 1 can be thought of as a signature of concentration of the turbulent transport on a subset of phase space with fractal geometry.  

\section{Fractional kinetic equations}

\subsection{Fractional diffusion equation}

The power-law dependence $D / \omega \propto Q^{\gamma}$ which we associate with the initial diffusion on a fractal cluster at percolation must have implications for the generalized form of the diffusion equation at time scales $a/u_\bot\ll t\lesssim\tau_c\ll 1/\omega$, for which the charge carriers move only on the fractal. Observe that the dispersion in Eq.~(\ref{9}) is sub-linear at these time scales because of the inverse power-law behavior $f(t/\tau_*)\propto (t/\tau_*)^{-\mu/\nu(2+\theta)}$ and that the function $f$ has not yet reached its asymptotic value. The sub-linear dispersion is explained by the trapping effect caused by cycles and dead-ends of the fractal (i.e., the ``nodes-links-blobs" model) (Ref. \cite{Naka} and Fig.~1 therein). In general, obstacles and traps act in a way as to introduce memory into the motion. One possible way to include the memory is to generalize Fick's first law to (where, to simplify the notations, we denote the position vector simply by ${\bf{r}}$)
\begin{equation}
\textbf{j} (t, {\bf{r}}) = - \int _{0}^{t} \Lambda (t - t^{\prime}) \nabla\rho (t^{\prime}, {\bf{r}}) dt^{\prime}\label{Fick} 
\end{equation}
with the continuity condition written as 
\begin{equation}
\frac{\partial}{\partial t}\rho (t, {\bf{r}}) = \int _{0}^{t} \Lambda (t - t^{\prime}) \nabla^2\rho (t^{\prime}, {\bf{r}}) dt^{\prime}.\label{Cont} 
\end{equation}
In the above, $\textbf{j} (t, {\bf{r}})$ is the probability current, which flows against the concentration gradient, and $\Lambda (t - t^{\prime})$ is a memory function, which is nonzero for time lags, for which the dynamics are influenced by the fractal geometric properties, and is identically zero or vanishing otherwise. In order to correctly introduce the memory function, we define first a generalized $f$ function via 
\begin{equation}
\hat f (s) = \cases{0 & if\, $s \leq 0$\cr f (s) & if\, $s > 0$}.\label{Cases} 
\end{equation}
Thus, by its definition, $\hat f(s)$ coincides with the $f(s)$ function for $s > 0$ and is identically zero for $s \leq 0$. Without loss in generality, we can set the asymptotic value of $f (s)$ to 1, thus exposing the definition of $\hat f(s)$ as follows:
\begin{equation}
\hat f (s) = \cases{0 & if\, $s \leq 0$\cr s^{\gamma - 1} / \Gamma (\gamma) & if\, $0 < s\ll s_c$\cr 1 & if\, $s\gg s_c$},\label{Cases2} 
\end{equation}
where $s_c\propto \tau_c$ is a crossover time scale, where the behavior changes from a power-law to the asymptotic constant value, and the gamma-function $\Gamma (\gamma)$ is introduced for the normalization reasons. Mathematically, the $\hat f (s)$ function offers a suitable generalization of the Heaviside step function by incorporating an intermediate, power-law behavior for $0 < s\ll s_c$. The Heaviside function is recovered from Eq.~(\ref{Cases2}) in the limit $\gamma\rightarrow 1$. Observe the following properties of the $\hat f(s)$ function: (i) $\hat f(s)$ is discontinuous in the origin for all $\gamma \leq 1$; and (ii) the integral $\int \hat f(s) ds$ converges for $s\rightarrow +0$ when $\gamma > 0$. Likewise the Heaviside step function and Dirac's delta function, the $\hat f (s)$ function and its derivatives must be considered as generalized functions. Now the memory function is defined as the time derivative of the $\hat f(s)$ function: 
\begin{equation}
\Lambda (s) = \frac{d}{ds}\hat f (s).\label{19} 
\end{equation}
We have, by means of Eq.~(\ref{19}), connected the memory function with the generalized scaling function $\hat f (s)$ in the random walk model. Clearly, $\Lambda (s)\rightarrow 0$ for $s\gg s_c$ and $\Lambda (s) \equiv 0$ for $s<0$. The latter condition is also required by causality. When $\gamma\rightarrow 1$, the $\Lambda (s)$ function coincides with Dirac's delta function: $\Lambda (s) = \delta (s)$. In this limit, the relation between $\textbf{j} (t, {\bf{r}})$ and $\nabla\rho (t, {\bf{r}})$ is local in time [see Eq.~(\ref{Fick})]. Accordingly, the dynamics are memoryless, corresponding to an ordinary Fickian diffusion and the frequency-independent asymptotic diffusion coefficient. Combining Eqs.~(\ref{Cont}) and~(\ref{19}) we have
\begin{equation}
\frac{\partial}{\partial t}\rho (t, {\bf{r}}) = \int _{0}^{t} \frac{d}{dt}\hat f (t - t^{\prime}) \nabla^2\rho (t^{\prime}, {\bf{r}}) dt^{\prime}\label{20} 
\end{equation}
or
\begin{equation}
\frac{\partial}{\partial t}\rho (t, {\bf{r}}) = -\int _{0}^{t} \frac{d}{dt^{\prime}}\hat f (t - t^{\prime}) \nabla^2\rho (t^{\prime}, {\bf{r}}) dt^{\prime},\label{21} 
\end{equation}
where the time derivative $d / dt^{\prime}$ acts only on $\hat f (t-t^{\prime})$. Integrating by parts in Eq.~(\ref{21}) with the initial condition $\nabla^2 \rho (t, {\bf{r}}) = 0$ for $t=0$ it is found that 
\begin{equation}
\frac{\partial}{\partial t}\rho (t, {\bf{r}}) = \int _{0}^{t} \hat f (t - t^{\prime}) \frac{\partial}{\partial t^{\prime}}\nabla^2\rho (t^{\prime}, {\bf{r}}) dt^{\prime},\label{22} 
\end{equation}
where use has been made of $\hat f(0) = 0$. Focusing on the intermediate, self-similar range, we take $\hat f(s) = s^{\gamma - 1} / \Gamma (\gamma)$, making it possible to rewrite Eq.~(\ref{22}) as 
\begin{equation}
\frac{\partial}{\partial t}\rho (t, {\bf{r}}) = \frac{1}{\Gamma (\gamma)}\int _{0}^{t} \frac{dt^{\prime}}{(t - t^{\prime})^{1-\gamma}} \frac{\partial}{\partial t^{\prime}}\nabla^2\rho (t^{\prime}, {\bf{r}}).\label{23} 
\end{equation}
The operator on the right hand side of Eq.~(\ref{23}) acting on $\nabla^2\rho (t^{\prime}, {\bf{r}})$ is known as the Caputo fractional derivative of order $0 < 1-\gamma < 1$, ${_0^c}D_t^{1-\gamma}$ (Ref. \cite{Podlubny} for details). With this last definition, Eq.~(\ref{23}) takes the form of a fractional diffusion equation 
\begin{equation}
\frac{\partial}{\partial t}\rho (t, {\bf{r}}) = {_0^c}D_t^{1-\gamma} \nabla^2\rho (t, {\bf{r}}).\label{24} 
\end{equation}
For $\gamma\rightarrow 1$, the Caputo derivative acts as a unit operator, thus yielding the familiar $-$ ``integer" $-$ diffusion equation with no integro-differentiation added. The same integer equation is obtained by identifying $\hat f(s)$ with the Heaviside step function in Eq.~(\ref{22}).

The end result of the discussion above is that the turbulent transport associated with a simple random walk process on a fractal cluster at percolation is described by a fractional time diffusion equation with the fractional derivative in the Caputo sense. This equation is different from the equation considered in Refs. \cite{Klafter,Sokolov,Rest} in that it uses the Caputo fractional derivative, ${_0^c}D_t^{1-\gamma}$ (instead of more familiar the so-called Riemann-Liouville fractional derivative \cite{Oldham}). 

\subsection{Fractional relaxation equation}

The characteristic function of the fractional diffusion equation, Eq.~(\ref{24}), obeys the fractional relaxation equation with the Caputo time derivative 
\begin{equation}
\frac{\partial}{\partial t}\rho (t, {\bf{k}}) = -{\bf k}^2{_0^c}D_t^{1-\gamma}\rho (t, {\bf{k}}).\label{Relax} 
\end{equation}
Initial-time behavior of $\rho (t, {\bf{k}})$ can be obtained by reducing the fractional derivative to   
\begin{equation}
{_0^c}D_t^{1-\gamma}\rho (t, {\bf{k}}) \simeq \frac{1}{\Gamma (\gamma)} t^{\gamma - 1}\rho (t, {\bf{k}}).\label{26} 
\end{equation}
Substituting into Eq.~(\ref{Relax}) and carrying out the trivial integration over $t$ leads to a stretched exponential form for the $\rho (t, {\bf{k}})$ function (see, also, Ref. \cite{Klafter} and Eq.~(B.7) therein):   
\begin{equation}
\rho (t, {\bf{k}}) \simeq \exp \{-{\bf k}^2 t^\gamma / \Gamma(\gamma + 1)\}.\label{KWW} 
\end{equation}
This stretched exponential relaxation describes the decay of charge density inhomogeneities in self-similar geometry. To obtain a 2D specific expression we set $\gamma = 2/3$ in Eq.~(\ref{KWW}), yielding $\rho (t, {\bf{k}}) \simeq \exp \{-{\bf k}^2 t^{2/3} / \Gamma(5/3)\}$. Apart from the plasma physics application discussed above we expect this relaxation pattern to also characterize charge relaxation in thin films of disordered solid materials. Some discussion of these properties can be found in Refs. \cite{PRB07,PLA,PRB01}.  

The main conclusion to be drawn from the above analysis is that the anomalous diffusion associated with the random walks at percolation has non-Markovian character at intermediate time scales that are in the range $a/u_\bot\ll t\lesssim\tau_c\ll 1/\omega$. For $t\gg \tau_c$, the fractal correlations, included in ${_0^c}D_t^{1-\gamma}$, vanish. Consistent with this loss of correlation the fractional diffusion and relaxation equations discussed above cross over to their integer-derivative counterparts as time $t\rightarrow\infty$. It is this loss of correlation that permits one to speak about the asymptotic transport process in terms of ``diffusion," in the true sense of the wording, and to obtain the ``anomalous" scaling law in Eq.~(\ref{10}) by limiting the power-law range of the $f (s)$ variation.  
    
\section{Including inertia effects}

The next contribution to the theory of turbulent diffusion involves the effects of particle inertia in the slowly varying electric and magnetic fields. Here, we consider a simplified model, which assumes the magnetic field to be uniform in space and time: ${\bf B} = {\rm const.}$ This model captures the essential physics due to inertia. Further generalization allowing the magnetic field to also be a function of time is basically obvious. In a constant ${\bf B}$ field with a slowly varying perpendicular electric field the cross-field guiding center drift velocity is accurately approximated by (top sign for electrons) \cite{Jeffrey}
\begin{equation}
{\bf u}_\bot (t) = \frac{{\bf E}_\bot\times{\bf B}}{B^2}\mp\frac{1}{\omega_c B}\frac{d{\bf E}_\bot}{dt},\label{PD} 
\end{equation}
where the last term accounts for the polarization drift and the time variation of electric field is assumed to be slow compared to the gyro frequency. Assuming a periodic time dependence ${\bf E}_\bot \propto e^{i\omega t}$ with the characteristic wave frequency $\omega \ll \omega_c$ one finds the velocity components to be 
\begin{equation}
u_x = E_y \mp (i\omega / \omega_c)E_x, 
\label{PD21} 
\end{equation}
\begin{equation}
u_y = -E_x \mp (i\omega / \omega_c)E_y,
\label{PD22} 
\end{equation}
where $B=1$ for simplicity. Eliminating $E_x$ in Eq.~(\ref{PD21}) yields 
\begin{equation}
u_x \mp (i\omega / \omega_c) u_y = E_y - (\omega / \omega_c)^2 E_y \approx E_y.
\label{PD31} 
\end{equation}
Likewise in Eq.~(\ref{PD22}): 
\begin{equation}
u_y \pm (i\omega / \omega_c) u_x = - E_x + (\omega / \omega_c)^2 E_x \approx -E_x.
\label{PD32} 
\end{equation}
Last terms on the right hand side of Eqs.~(\ref{PD31}) and~(\ref{PD32}) could be neglected since they involve the same components of electric field as the leading terms and are smaller by $(\omega / \omega_c)^2$. Applying $\partial / \partial x$ to Eq.~(\ref{PD31}); then $\partial / \partial y$ to Eq.~(\ref{PD32}); and adding the two equations together one finds
\begin{equation}
\frac{\partial u_x}{\partial x} + \frac{\partial u_y}{\partial y} = \pm \frac{i\omega}{\omega_c} \left(\frac{\partial u_y}{\partial x} - \frac{\partial u_x}{\partial y}\right).
\label{Comp} 
\end{equation}
Here, use has been made of ${\bf E}_\bot = -\nabla\Phi$. Equation~(\ref{Comp}) shows that the polarization drift introduces subtle compressibility into the motion. Clearly, the compressibility effect scales linearly with the mass-to-charge ratio (via the dependence on the cyclotron frequency). Topologically, the compressibility of advecting flow gives rise to limit cycles and stable foci, which attract and trap tracer particles (Ref. \cite{Isi} and Fig.~30 therein). Thus, we expect the turbulent diffusivities of inertial tracers to be comparatively smaller than the diffusivities of ideal particles due to compression by the polarization drift. Equations~(\ref{PD31}) and~(\ref{PD32}) suggest the compression effect be described in terms of the effective velocity flow, with the components of the corresponding velocity vector defined as $\tilde u_x = u_x \mp (i\omega / \omega_c) u_y$ and $\tilde u_y = u_y \pm (i\omega / \omega_c) u_x$, and the characteristic flow speed $\tilde u_\bot = \sqrt{\tilde u_x^2 + \tilde u_y^2} \simeq u_\bot\sqrt{1-\omega^2 /\omega_c^2}$. With these definitions one introduces the effective Kubo number by means of $\tilde Q \simeq \tilde u_\bot / \omega\xi_\bot$. That is,
\begin{equation}
\tilde Q \simeq Q \sqrt{1-\frac{\omega^2}{\omega_c^2}}.
\label{CV} 
\end{equation}
Accordingly, the effective diffusion coefficient is obtained from Eq.~(\ref{11}), in which one uses $\tilde Q$ instead of $Q$. Remembering that $\gamma = 2/3$ at percolation, and keeping the first non-vanishing correction due to inertia, we find
\begin{equation}
D\simeq (\ln K)^{1/3} (a/2\xi_\bot)^{2/3} \omega \xi_\bot^2 (1-\omega^2 / 3\omega_c^2) Q^{2/3}. \label{Inertia} 
\end{equation}
Prospective applications of the results obtained can be proposed for the impurity transport in fusion plasma. Recently, the turbulent transport of inertial impurities have been studied by numerical simulations in Ref. \cite{Impur} and the specific physics consequences arising from compressibility have been placed in the center of attention.

\section{Overall Summary and Final Remarks}

In the present work, we have exposed a few crucial physics issues behind the so-called ``percolation" transport in low-frequency, electrostatic (anisotropic magnetic) turbulence, basing our investigations on the formalism of Hamiltonian dynamics, random walk models, and fractional derivative equations. The central problem being addressed is the scaling of the turbulent diffusion coefficient with the fluctuation strength in the limit of slow fluctuation frequencies (large Kubo numbers). In this limit, the transport is found to exhibit pseudochaotic, rather than simply chaotic, properties. ``Pseudochaotic" means random non-chaotic dynamics with zero Lyapunov exponents \cite{Report,JMPB,PD2004}. In our description, pseudochaos occurs as a consequence of the concentration of the resonant motions on a subset of phase space with fractal geometry. Because of the strongly reduced phase space, the dynamics are characterized by anomalously slow mixing properties associated with the vanishing Kolmogorov-Sinai entropy. 

As a simplified, microscopic model we considered a random walk model on a fractal cluster at percolation (``the unbiased ant in the fractal labyrinth") \cite{Gennes}. Based on this model, we found the low-frequency, percolation scaling of the turbulent diffusion coefficient to be given by $D/\omega \propto Q^\gamma$ with $\gamma = 2/3$ (here $Q\gg 1$ is the Kubo number), in agreement with the work in Ref. \cite{PRE01} and at variance with the prediction $\gamma = 7/10$ in Refs. \cite{Kalda,Kalda2,Isi}. When the non-fractal limit is taken in the model (i.e., the random motions are thought of as space-filling),  Bohm scaling with $\gamma = 1$ is reproduced. In this spirit, a deviation from Bohm scaling, associated with the $\gamma$ exponent smaller than 1, can be interpreted as a signature of concentration of the transport processes on a fractal geometry.  

Focusing on the non-Markovian properties of the transport, by introducing a generalized memory function for the random walks at percolation, we have derived a fractional diffusion equation with the time derivative in the Caputo sense (as opposed to a more conventional definition of the fractional differentiation in the Riemann-Liouville sense). In our study the non-Markovian property has occurred as a consequence of fractality and, therefore, has had a purely geometric origin.     

Finally, we have discussed a simple generalization of the model treatment described above by taking into account finite particle inertia. Inertia enters the model equations in the form of the polarization drift and leads to a decreased particle diffusivity because of compression effects. These results may find further application in describing the impurity transport in fusion plasma. 
 
We have, in the present work, significantly simplified the presentations by assuming a characteristic microscopic time and spatial scales of the random motion. Extensions to include a distribution of time intervals between consecutive steps are straightforward and have, in the case of homogeneous support, discussed in Refs. \cite{Sokolov,Klafter} on the basis of continuous time random walks (CTRWs) \cite{CTRW}. In general we expect the effect of time scale distribution to slow down the anomalous diffusion due to fractal geometry. In this respect, a fractional $\gamma$ smaller than 2/3 may be conceivable. We remark, however, that, while the detailed, microscopic picture of the random motion may vary, we expect the basic physical properties discussed here to remain essentially the same.    

CTRW-like models and their derivatives are further generalized to include a distribution of jump lengths \cite{Klafter,Uchaikin,Compte,Chechkin,Rest} physically corresponding to nonlocal transport in phase space. On the level of CTRWs, fractional diffusion models of perturbative transport in magnetically confined fusion plasma including nonlocal transport have been obtained in Refs. \cite{Castillo1,Castillo2,Sanchez,Castillo3}, where one also finds a discussion of the numerical simulation results. 

There can be various physical mechanisms at play to give rise to nonlocal transport. One such mechanism can be associated with mode coupling and build-up of correlations [not included in the wave-like Hamiltonian in Eq.~(\ref{5})] leading to the formation of large-scale coherent structures in the turbulent flow. It is found in direct numerical simulations of forced and dissipative turbulence that the presence of coherent structures leads to a spatially nonuniform transport \cite{Provenzale}. Consistent with this property, in electrostatic drift-wave turbulence, coherent, vortex structures are found to enhance the diffusion in the direction of the background density gradient \cite{Naulin,Naulin2}. The propagation of coherent structures can also cause significant broadening of the turbulent region and in this fashion affect the scaling properties of the transport \cite{White,Guo}. In burning plasmas, where the energetic ions (MeV energies) and charged fusion products constitute a significant fraction of the total plasma energy density, the coupling between the nonlinear energetic particle modes \cite{EPM}, mediated by the energetic particles themselves, results in the transition to strong, convective-like transport by radially amplifying avalanches \cite{Zonca95,ZoncaChen}. The process obeys complex nonlinear parabolic equation, which, under some nonrestrictive assumptions regarding the shape of the energetic particle source function, can be cast \cite{Zonca} in the generic form of a fractional nonlinear Schr\"odinger equation \cite{UFN}. Indeed the latter equation, which is closely related with the fractional Ginzburg-Landau equation \cite{PLA05}, describes the fractional dynamics of coupled nonlinear oscillators with long-range interaction \cite{Tarasov}. After all, we address nonlocal edge phenomena in magnetically confined fusion plasma as for instance the problem of anomalously fast response in the plasma core to a cold pulse edge perturbation \cite{Castillo3,EPS06}. Analyses of these general phenomena remains to be carried out.


{\bf Acknowledgments:} Illuminating discussions with J. Juul Rasmussen, L. M. Zelenyi, G. Zimbardo, and F. Zonca on various topics addressed in this work are gratefully acknowledged. This study was partially supported by INTAS grant 06-1000017-8943. 


\begin{thebibliography}{}

\bibitem{Kalda}
A. V. Gruzinov, M. B. Isichenko, and Ya. L. Kalda, Zh. Eksp. Teor. Fiz. {\bf 97}, 476 (1990) [Sov. Phys. JETP {\bf 70}, 263 (1990)]. 

\bibitem{Kalda2}
M. B. Isichenko, Plasma Phys. Control. Fusion {\bf 33}, 809 (1991).

\bibitem{Horton}
M. B. Isichenko and W. Horton, Comments Plasma Phys. Control. Fusion {\bf 14}, 249 (1991). 

\bibitem{Isi}
M. B. Isichenko, Rev. Mod. Phys. {\bf 64}, 961 (1992).

\bibitem{Misguich}
J.-D. Reuss and J. H. Misguich, Phys. Rev. E {\bf 54}, 1857 (1996).

\bibitem{Vlad}
M. Vlad, F. Spineanu, J. H. Misguich, and R. Balescu, Phys. Rev. E {\bf 58}, 7359 (1998).

\bibitem{Vlad2}
J.-D. Reuss, M. Vlad, and J. H. Misguich, Phys. Lett. A {\bf 241}, 94 (1998).

\bibitem{Zim00}
G. Zimbardo, P. Veltri, and P. Pommois, Phys. Rev. E {\bf 61}, 1940 (2000).

\bibitem{Zim01}
P. Pommois, P. Veltri, and G. Zimbardo, Phys. Rev. E {\bf 63}, 066405 (2001).

\bibitem{Veltri}
P. Pommois, G. Zimbardo, and P. Veltri, Phys. Plasmas {\bf 14}, 012311 (2007).

\bibitem{Bohm}
T. H. Dupree, Phys. Fluids {\bf 10}, 1049 (1967).

\bibitem{PRE01}
A. V. Milovanov, Phys. Rev. E {\bf 63}, 047301 (2001).

\bibitem{PScripta}
F. Chiaravalloti, A. V. Milovanov, and G. Zimbardo, Phys. Scr. {\bf T122}, 79 (2006).

\bibitem{Chaos}
G. M. Zaslavsky, Chaos {\bf 4}, 25 (1994).

\bibitem{PhysicaD}
G. M. Zaslavsky, Physica D {\bf 76}, 110 (1994).

\bibitem{Saichev}
A. I. Saichev and G. M. Zaslavsky, Chaos {\bf 7} (4), 753 (1997).

\bibitem{Report}
G. M. Zaslavsky, Phys. Rep. {\bf 371},  461 (2002).

\bibitem{JMPB}
O. Lyubomudrov, M. Edelman, and G. M. Zaslavsky, Intl. J. Modern Phys. B {\bf 17}, 4149 (2003).  

\bibitem{PD2004}
G. M. Zaslavsky and M. A. Edelman, Physica D {\bf 193}, 128 (2004).

\bibitem{Podlubny}
I. Podlubny, {\it Fractional Differential Equations} (Academic Press, San Diego, 1999). 

\bibitem{Oldham}
K. B. Oldham and J. Spanier, {\it The Fractional Calculus} (Academic Press, San Diego, Calif. 1974).

\bibitem{Klafter}
R. Metzler and J. Klafter, Phys. Rep. {\bf 339}, 1 (2000).

\bibitem{Sokolov}
I. M. Sokolov, J. Klafter, and A. Blumen, Phys. Today {\bf 55}, 48 (2002).

\bibitem{Rest}
R. Metzler and J. Klafter, J. Phys. A: Math. Gen. {\bf 37}, R161 (2004).

\bibitem{Uchaikin}
V. V. Uchaikin, Phys. Usp. {\bf 46}, 821 (2003).

\bibitem{Galeev}
A. A. Galeev, M. M. Kuznetsova, and L. M. Zelenyi, Space Sci. Rev. {\bf 44}, 1 (1986).

\bibitem{UFN}
L. M. Zelenyi and A. V. Milovanov, Phys. Usp. {\bf 47}, 749 (2004). 

\bibitem{Zonca}
F. Zonca, S. Briguglio, L. Chen, G. Fogaccia, T. S. Hahm, A. V. Milovanov, and G. Vlad, Plasma Phys. Controlled Fusion {\bf 48}, B15 (2006).

\bibitem{Zonca2}
F. Zonca, Intl. J. Modern Phys. A {\bf 23}, 1165 (2008).  

\bibitem{Taylor}
M. N. Rosenbluth, R. Z. Sagdeev, J. B. Taylor, and G. M. Zaslavsky, Nucl. Fusion {\bf 6}, 297 (1966).

\bibitem{Naulin}
V. Naulin, A. H. Nielsen, and J. Juul Rasmussen, Phys. Plasmas {\bf 6}, 4575 (1999).

\bibitem{Horton2}
W. Horton, Rev. Mod. Phys. {\bf 71}, 735 (1999).

\bibitem{Bitten}
J. A. Bittencourt, {\it Fundamentals of Plasma Physics} (Pergamon Press, New York, 1986).

\bibitem{Jeffrey}
J. Freidberg, {\it Plasma Physics and Fusion Energy} (Cambridge Univ. Press, Cambridge, 2007).

\bibitem{Feder}
J. Feder, {\it Fractals} (Plenum, New York, 1988). 

\bibitem{LeMehaute}
A. Le Mehaute, {\it Fractal Geometries: Theory and Applications} (CRC Press, Boca Raton, FL, 1991).

\bibitem{Burlaga86}
L. F. Burlaga and L. W. Klein, J. Geophys. Res. {\bf 91}, 347 (1986). 

\bibitem{JGR96}
A. V. Milovanov, L. M. Zelenyi, and G. Zimbardo, J. Geophys. Res. {\bf 101}, 19 903 (1996). 

\bibitem{Borovsky}
J. M. Weygand, M. G. Kivelson, K. K. Khurana, H. K. Schwarzl, S. M. Thompson, R. L. McPherron, A. Balogh, L. M. Kistler, M. L. Goldstein, J. Borovsky, and D. A. Roberts, J. Geophys. Res. {\bf 110}, A01205 (2005).

\bibitem{Carreras}
B. A. Carreras, B. van Milligen, C. Hidalgo, R. Balbin, E. Sanchez, I. Garcia-Cortes, M. A. Pedrosa, J. Bleuel, and M. Endler, Phys. Rev. Lett. {\bf 83}, 3653 (1999). 

\bibitem{Zaslav}
G. M. Zaslavsky, M. Edelman, H. Weitzner, B. Carreras, G. McKee, R. Bravenec, and R. Fonck, Phys. Plasmas {\bf 7}, 3691 (2000). 

\bibitem{Siam}
B. B. Mandelbrot and J. Van Ness, SIAM Rev. {\bf 10}, 422 (1968).

\bibitem{Affine}
B. B. Mandelbrot, {\it Gaussian Self-Affinity and Fractals} (Springer-Verlag, Berlin, 2002).

\bibitem{PRE00}
A. V. Milovanov and G. Zimbardo, Phys. Rev. E {\bf 62}, 250 (2000).

\bibitem{Zaslavsky}
G. M. Zaslavsky, {\it Statistical Irreversibility in Nonlinear Systems} (Nauka, Moscow, 1970).

\bibitem{ZaslavskyUFN}
G. M. Zaslavsky and B. V. Chirikov, Phys. Usp. {\bf 14}, 549 (1972). 

\bibitem{Sagdeev}
G. M. Zaslavsky and R. Z. Sagdeev, {\it Introduction to the Nonlinear Physics. From Pendulum to Turbulence and Chaos} (Nauka, Moscow, 1988).

\bibitem{Aharony}
T. Grossman and A. Aharony, J. Phys. A {\bf 19}, L745 (1986).

\bibitem{Unpublished}
A. V. Milovanov, R. Bitane, and G. Zimbardo, submitted to Plasma Phys. Control. Fusion. 

\bibitem{Zimbardo}
G. Zimbardo, R. Bitane, P. Pommois, and P. Veltri, Plasma Phys. Control. Fusion {\bf 51}, 015005 (2009).

\bibitem{Gefen}
Y. Gefen, A. Aharony, and S. Alexander, Phys. Rev. Lett. {\bf 50}, 77 (1983).

\bibitem{Straley}
J. P. Straley, J. Phys. C {\bf 13}, 2991 (1980).

\bibitem{Gennes}
P. G. de Gennes, Recherche {\bf 7}, 919 (1976).

\bibitem{Havlin}
S. Havlin and D. ben-Avraham, Adv. Phys. {\bf 51}, 187 (2002).

\bibitem{CTRW}
E. W. Montroll and G. H. Weiss, J. Math. Phys. {\bf 10}, 753 (1969).

\bibitem{Stauffer}
D. Stauffer, Phys. Rep. {\bf 54}, 1 (1979).

\bibitem{Naka}
T. Nakayama, K. Yakubo, and R. L. Orbach, Rev. Mod. Phys. {\bf 66}, 381 (1994).

\bibitem{AO}
S. Alexander and R. Orbach, J. Phys. Lett. Paris {\bf 43}, L625 (1982).

\bibitem{Bouchaud}
J.-P. Bouchaud and A. Georges, Phys. Rep. {\bf 195}, 127 (1990). 

\bibitem{Coniglio}
A. Coniglio, J. Phys. A {\bf 15}, 3829 (1982).

\bibitem{Havlin2}
D. ben-Avraham and S. Havlin, {\it Diffusion and Reactions in Fractals and Disordered Systems} (Cambridge University Press, Cambridge, 2000). 

\bibitem{PRE97}
A. V. Milovanov, Phys. Rev. E {\bf 56}, 2437 (1997).

\bibitem{Stat}
J. M. Normand, H. J. Herrmann, and M. Hajjar, J. Stat. Phys. {\bf 52}, 441 (1988).

\bibitem{PRB07}
A. V. Milovanov, K. Rypdal, and J. Juul Rasmussen, Phys. Rev. B {\bf 76}, 104201 (2007). 

\bibitem{PLA}
A. V. Milovanov, J. Juul Rasmussen, and K. Rypdal, Phys. Lett. A {\bf 372}, 2148 (2008).

\bibitem{PRB01}
A. V. Milovanov and J. Juul Rasmussen, Phys. Rev. B {\bf 64}, 212203 (2001); ibid. {\bf 66}, 134505 (2002). 

\bibitem{Impur}
M. Priego, O. E. Garcia, V. Naulin, and J. Juul Rasmussen, Phys. Plasmas {\bf 12}, 062312 (2005).

\bibitem{Compte}
A. Compte, Phys. Rev. E {\bf 53}, 4191 (1996).

\bibitem{Chechkin}
R. Metzler, A. V. Chechkin, V. Yu. Gonchar, and J. Klafter, Chaos, Solitons, and Fractals {\bf 34}, 129 (2007).

\bibitem{Castillo1}
D. del-Castillo-Negrete, B. A. Carreras, and V. E. Lynch, Phys. Plasmas {\bf 11}, 3854 (2004).

\bibitem{Castillo2}
D. del-Castillo-Negrete, Phys. Plasmas {\bf 13}, 082308 (2006).

\bibitem{Sanchez}
B. A. Carreras, V. E. Lynch, B. Ph. van Milligen, and R. S\'anchez, Phys. Plasmas {\bf 13}, 062301 (2006).

\bibitem{Castillo3}
D. del-Castillo-Negrete, P. Mantica, V. Naulin, J. Juul Rasmussen, and JET EFDA contributors, Nucl. Fusion {\bf 48}, 075009 (2008).

\bibitem{Provenzale}
A. Babiano and A. Provenzale, J. Fluid Mech. {\bf 574}, 429 (2007).

\bibitem{Naulin2}
V. Naulin, O. E. Garcia, A. H. Nielsen, and J. Juul Rasmussen, Phys. Lett. A {\bf 321}, 355 (2004).

\bibitem{White}
L. Chen, R. B. White, and F. Zonca, Phys. Rev. Lett. {\bf 92}, 075004 (2004).

\bibitem{Guo}
Z. Guo, L. Chen, and F. Zonca, arXiv:0902.3752v1 [physics.plasm-ph] (submitted to Phys. Rev. Lett.) 

\bibitem{EPM}
L. Chen, Phys. Plasmas {\bf 1} (5), 1519 (1994).

\bibitem{Zonca95}
F. Zonca, S. Briguglio, L. Chen, G. Fogaccia, and G. Vlad, Nucl. Fusion {\bf 45}, 477 (2005).

\bibitem{ZoncaChen}
L. Chen and F. Zonca, Nucl. Fusion {\bf 47}, S727 (2007).

\bibitem{PLA05}
A. V. Milovanov and J. Juul Rasmussen, Phys. Lett. A {\bf 337}, 75 (2005).

\bibitem{Tarasov}
V. E. Tarasov and G. M. Zaslavsky, Chaos {\bf 16}, 023110 (2006).

\bibitem{EPS06}
J. Juul Rasmussen, V. Naulin, J. S. L\"onnroth, P. Mantica, and V. Parail, Europhys. Conference Abstracts {\bf 32B}, P1-076 (2006).  


\end{thebibliography}

\end{document}